\documentclass[reprint, amsmath, amssymb, aps, prb, superscriptaddress, nofootinbib]{revtex4-2}

\usepackage{graphicx}
\usepackage{dcolumn}
\usepackage{bm}
\usepackage{mhchem}
\usepackage{hyperref}
\usepackage{mathrsfs}
\usepackage{physics}
\usepackage{xcolor}
\usepackage{txfonts}

\usepackage{color}

\usepackage{CJKutf8}

\usepackage[normalem]{ulem}
\newcommand{\MT}[1]{\textcolor{blue}{#1}}

\begin{document}

\preprint{CeRu2Si2_AHE}

\title{Impact of localized $4f$ magnetism on Berry curvature\\
       and the intrinsic anomalous Hall effect in \ce{CeRu2Si2}
}
\author{Takehiro Kobayashi}
\affiliation{Department of Materials Science, Graduate School of Engineering, Osaka Metropolitan University, Sakai, Osaka 599-8531, Japan}
\author{Shunsuke Hirai}
\affiliation{Department of Materials Science, Graduate School of Engineering, Osaka Metropolitan University, Sakai, Osaka 599-8531, Japan}
\author{Ibuki Terada}
\email{Corresponding author: iterada@omu.ac.com}
\affiliation{Department of Materials Science, Graduate School of Engineering, Osaka Metropolitan University, Sakai, Osaka 599-8531, Japan}
\author{Michi-To Suzuki}
\email{Contact author: mts@omu.ac.com}
\affiliation{Department of Materials Science, Graduate School of Engineering, Osaka Metropolitan University, Sakai, Osaka 599-8531, Japan}
\affiliation{Center for Spintronics Research Network, Graduate School of Engineering Science, Osaka University, Toyonaka, Osaka 560-8531, Japan}

\date{\today}

\begin{abstract}
The intrinsic anomalous Hall effect (AHE), originating from Berry curvature, has been extensively studied in $f$-electron systems, including \ce{Ce}-based compounds.
However, how the magnetism of localized $f$ electrons affect the intrinsic AHE remains unclear.
Here, we investigate this issue in \ce{CeRu2Si2} by calculating the intrinsic anomalous Hall conductivity (AHC) for several occupied \ce{Ce}-$4f$ states using first-principles LDA+$U$ calculations combined with Wannier interpolation method.
We find that the intrinsic AHC remains finite even when the \ce{Ce}-$4f$ states are localized and do not contribute directly to the Fermi surface, and that its magnitude depends on which \ce{Ce}-$4f$ state is occupied.
We show that localized $4f$-electron magnetism induces exchange splitting in the conduction bands, generating Berry curvature in specific regions near the Fermi surface.
This mechanism enables conduction bands that are derived from nonmagnetic orbitals and form the Fermi surface to contribute to the intrinsic AHE, suggesting a route to realizing large intrinsic AHE in $f$-electron materials.
\end{abstract}

\maketitle

\section{Introduction}
The anomalous Hall effect (AHE) is a transport phenomenon in which an electric current flows perpendicular to an applied electric field without an external magnetic field~\cite{Nagaosa2010AHE,Xiao2010BerryPhase}. 
Although this Hall response has been understood as a transport phenomenon specific to ferromagnetic metals, the observation of a large AHE in \ce{Mn3Sn} has expanded the scope of AHE research to $d$-electron antiferromagnets~\cite{SmejkalNatRevMater6-482-496,NakatsujiNat527,NayakSciAdv2,KiyoharaPRA5,Kotegawa_npjQuanMater8,AraiJPSJ93,doi:10.1126/sciadv.aaz8809,Feng2022}.
Recently, a large anomalous Hall conductivity (AHC) of $550~\mathrm{S/cm}$ has been observed in the $f$-electron antiferromagnet \ce{Ce2CuGe6}~\cite{PhysRevLett.133.106301}.
An important finding in Ref.~\cite{PhysRevLett.133.106301} is an intrinsic contribution of approximately $80~\mathrm{S/cm}$ to the AHC of \ce{Ce2CuGe6}.
This is counterintuitive because the localized magnetic \ce{Ce}-$4f$-electrons are not expected to contribute directly to the Fermi surface and hence to generate Berry curvature directly.
This raises the broader question of how localized $4f$-electrons affect the intrinsic AHE in $f$-electron magnetic systems, including ferromagnets.

To address this issue in a ferromagnetic setting, we focus on the Ce-based compound \ce{CeRu2Si2}, whose electronic structure has been extensively studied both experimentally and theoretically~\cite{Haen1987,BESNUS1985779,https://doi.org/10.1002/pssb.2221410226,Boucherle_2001,SuzukiJPSJ792}.
\ce{CeRu2Si2} crystallizes in the body-centered tetragonal \ce{ThCr2Si2}-type structure and is paramagnetic at zero magnetic field.
When an external magnetic field is applied, \ce{CeRu2Si2} undergoes a metamagnetic transition from the paramagnetic state to the ferromagnetic state~\cite{Haen1987}.
It has been experimentally established that the occupied \ce{Ce}-$4f$ state in ferromagnetic \ce{CeRu2Si2} is the $\ket{j=5/2,j_z=+5/2}$  state~\cite{BESNUS1985779,https://doi.org/10.1002/pssb.2221410226,Boucherle_2001}.
Here, $j$ denotes the total angular momentum resulting from the coupling of the orbital and spin angular momenta, and $j_z$ denotes its $z$ component.

\ce{CeRu2Si2} is particularly suitable for investigating the role of localized \ce{Ce}-$4f$ electrons in the intrinsic AHE because first-principles local density approximation (LDA)+$U$ calculations allow us to self-consistently obtain not only the occupied $\ket{j=5/2,j_z=+5/2}$ state but also other occupied $f$ states arising from the crystal-field splitting, as demonstrated below.
Taking advantage of this feature, we use \ce{CeRu2Si2} as a model system to investigate how different occupied $f$ states affect the AHC and to understand the underlying mechanism using first-principles LDA+$U$ calculations combined with the Wannier interpolation method.
We find that the AHC exhibits distinct values for different occupied $f$ states.
Our results indicate that these differences arise from changes in the Berry curvature distribution resulting from exchange splitting of the conduction bands induced by localized $f$-electron magnetism.

\section{Method}
We calculate the magnetic states of \ce{CeRu2Si2} using first-principles LDA+$U$ calculations as implemented in Quantum ESPRESSO, based on the Liechtenstein formulation~\cite{PhysRevB.52.R5467} and its extension including the spin-off-diagonal components of the on-site density matrix in the presence of spin–orbit coupling. 
A related LDA+$U$ treatment including spin-off-diagonal contributions was previously applied to \ce{CeRu2Si2}~\cite{SuzukiJPSJ792}.
In the following, we describe how different localized $f$ states are obtained through self-consistent calculations and how the anomalous Hall conductivity (AHC) is evaluated.

\subsection{LDA+$U$ method}
In the LDA+$U$ method, we consider the effective single-electron Hamiltonian, $H = H_{\mathrm{LDA}} + V_U[n^{il}_{mm'ss'}]$, 
where $H_{\mathrm{LDA}}$ denotes the Kohn-Sham Hamiltonian in the LDA. 
Here, $V_U$ denotes the LDA + $U$ correction term, which is a functional of the density matrix $n^{il}_{mm'ss'}$ and introduces an orbital-dependent on-site Coulomb interaction for the localized electrons with angular momentum $l$ on atom $i$.  
Here, $m$ and $s$ denote the magnetic quantum number and spin, respectively.
The density matrix $n^{il}_{mm'ss'}$ is defined by
\begin{equation}
n^{il}_{mm'ss'} =
\sum_{\bm kb}
\braket{\phi^{i}_{lms}}{\psi_{\bm kb}}
\braket{\psi_{\bm kb}}{\phi^{i}_{lm's'}},
\end{equation}
where $\phi^{i}_{lms}=\mathcal{Y}^{i}_{lm}\,\chi_s$ denotes a localized basis consisting of the real spherical harmonics $\mathcal{Y}^{i}_{lm}$ and the spin function $\chi_s$, and $\psi_{\bm kb}$ denotes the Bloch eigenstate with wave vector ${\bm k}$ and band index $b$.

In \ce{CeRu2Si2}, the \ce{Ce}-$4f$ states split into three doubly degenerate pairs, $\ket{j=5/2,j_z=\pm 1/2}$, $\ket{j=5/2,j_z=\pm 3/2}$, and $\ket{j=5/2,j_z=\pm 5/2}$, due to the combined effects of strong spin-orbit coupling and the tetragonal crystal field~\cite{SuzukiJPSJ792}.
The degeneracy of each pair is further lifted by the breaking of time-reversal symmetry.
Experimentally, the ferromagnetic state of \ce{CeRu2Si2} has been shown to be characterized by the occupation of the $\ket{j=5/2,j_z=+5/2}$ state~\cite{BESNUS1985779,https://doi.org/10.1002/pssb.2221410226,Boucherle_2001}.

In the LDA+$U$ calculations, we obtain self-consistent solutions corresponding not only to the experimentally realized $\ket{j=5/2, j_z=+5/2}$ state but also to other $4f$ occupation states, such as $\ket{j=5/2, j_z=+1/2}$ and $\ket{j=5/2, j_z=+3/2}$.
For each target $\ket{j,j_z}$ state, the initial density matrix is defined in
the $\ket{j,j_z}$ basis by setting the corresponding diagonal element to unity
and the other elements to zero.
The $\ket{j,j_z}$ states are related to the complex spherical-harmonic basis
through the Clebsch--Gordan coefficients,
\begin{equation}
    \ket{j,j_z}
    =
    \sum_{m,s}
    C^{j j_z}_{lms}
    \ket{l,m}\ket{s}.
\end{equation}
Using this transformation together with the transformation from complex to
real spherical harmonics, the initial density matrix is transformed to the
localized $\ket{\phi^i_{lms}}$ basis used in Quantum ESPRESSO.

\subsection{Anomalous Hall conductivity}
We analyze the intrinsic AHC of the ferromagnetic \ce{CeRu2Si2} using the Wannier interpolation method.
When the ferromagnetic moment of \ce{CeRu2Si2} is aligned along the $z$ direction, the $xy$ component of AHC tensor can be finite.
The intrinsic AHC is written as
\begin{equation}
\sigma_{xy}^{\mathrm{AHC}}
=
-\frac{e^{2}}{\hbar}
\int \frac{d^{3}k}{(2\pi)^{3}}
\sum_{b}
\Omega^z_{\bm kb}
f(\varepsilon_{\bm kb}-\mu),
\end{equation}
where $f(\varepsilon_{\bm kb}-\mu)$, $\varepsilon_{\bm kb}$, and $\mu$ denote the Fermi-Dirac distribution function, the energy band, and the chemical potential, respectively.
The $z$ component of the Berry curvature, $\Omega^z_{\bm kb}$ is expressed as
\begin{equation}
\Omega^z_{\bm kb}
=
-2\,\mathrm{Im}
\sum_{a(\neq b)}
\frac{
  \mel**{u_{\bm kb}}{\partial_{k_x} H_{\bm k}}{u_{\bm ka}}\mel**{u_{\bm ka}}{\partial_{k_y} H_{\bm k}}{u_{\bm kb}}
}{
(\varepsilon_{\bm ka}-\varepsilon_{\bm kb})^{2}
},
\end{equation}
where $\ket{u_{\bm ka}}$ denotes the eigenstate of the Bloch Hamiltonian $H_{\bm k}$~\cite{Nagaosa2010AHE,PhysRevB.74.195118}.

\subsection{Calculation Details}
We calculate the electronic structure of the ferromagnetic \ce{CeRu2Si2} using Quantum ESPRESSO package~\cite{ref9}.
We use the experimental lattice constants $a=4.196~\AA$ and $c=9.797~\AA$~\cite{ref8}.
The calculations were performed within the LDA+$U$ method including the spin-orbit coupling.
Ultrasoft pseudopotentials were employed, and the cutoff energies for plane-wave basis set and charge density were set to $200~\rm{Ry}$ and $2000~\rm{Ry}$, respectively.
The $k$-point meshes in Brillouin zone were set to $8\times 8\times 8$ for the self-consistent-field calculations and $16\times 16\times 16$ for the calculation of the density of states (DOS).

We constructed a tight-binding model for ferromagnetic \ce{CeRu2Si2} from the first-principles electronic structure and calculated the AHC using the Wannier90 package~\cite{Pizzi_2020}.
We used the $s$, $p$, $d$, and $f$ orbitals of \ce{Ce}, the $s$, $p$, and $d$ orbitals of \ce{Ru}, and the $s$ and $p$ orbitals of \ce{Si} as projection functions for the tight-binding model.
The $k$-point mesh for the AHC calculations was $80\times80\times80$, for which we confirmed that the AHC was well converged.

\section{Results}

\subsection{Electronic structure}
\begin{figure}[t]
  \centering
  \includegraphics[width=\columnwidth]{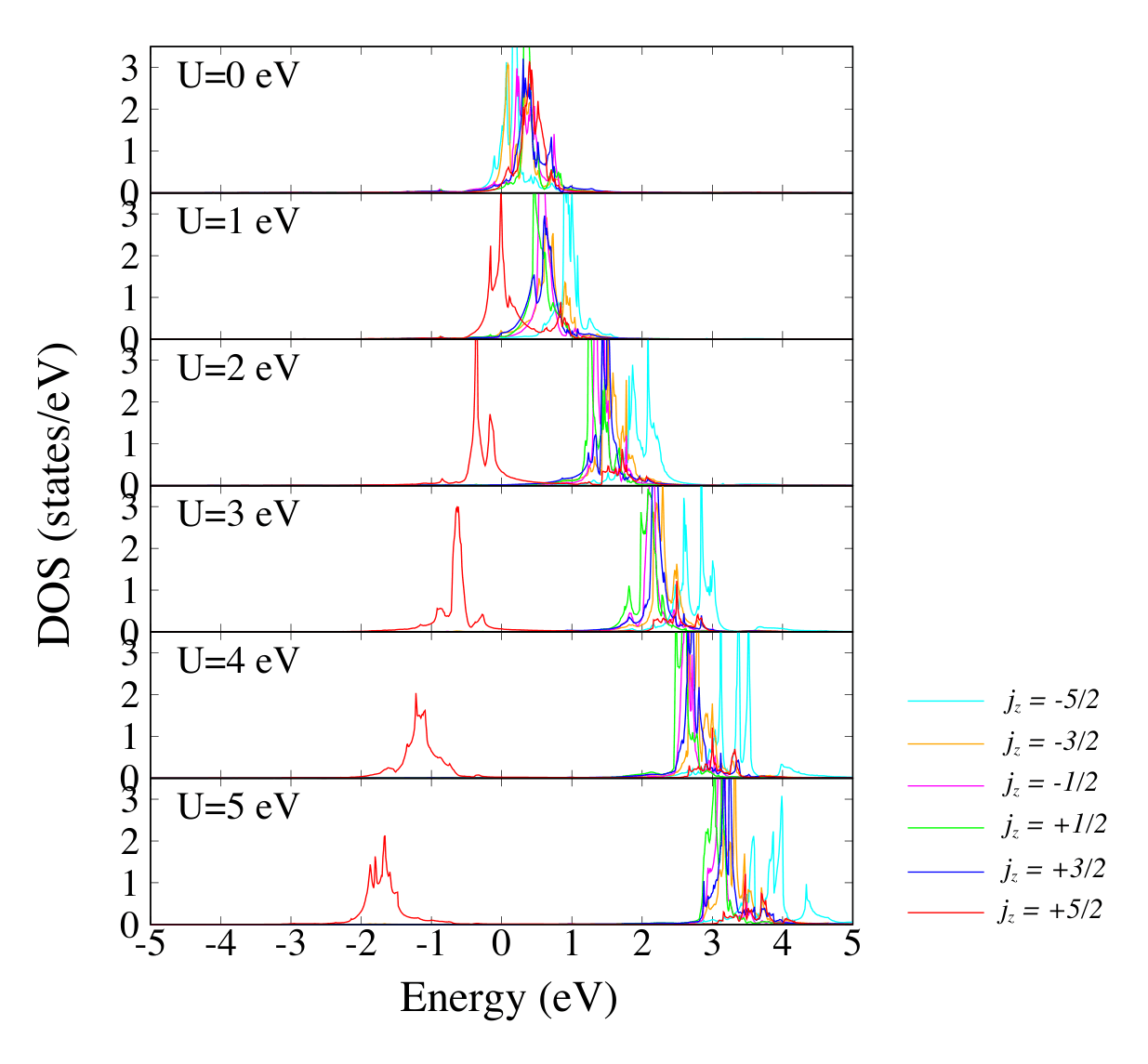}
  \caption{$j_z$-resolved DOS for the $j=5/2$ orbitals in the LDA+$U$ calculations. The $\ket{j=5/2,\, j_z=+5/2}$ orbital is used as the initial occupation. The Fermi level is set to $0~\mathrm{eV}$. 
  }
  \label{fig:fig1}
\end{figure}

We first examine the localization of the \ce{Ce}-$4f$ states with increasing $U$ through self-consistent calculations starting from an initial occupation corresponding to the $\ket{j=5/2,j_z=+5/2}$ state.
Figure~\ref{fig:fig1} represents the $j_z$-resolved DOS of the \ce{Ce}-$4f$ states projected onto the $j=5/2$ subspace for different values of $U$.
At $U=0$, corresponding to the LDA calculation, the \ce{Ce}-$4f$ states are located near the Fermi level and directly contribute to the energy band near the Fermi level.
As $U$ increases, the $j=5/2$ states separate into the occupied $\ket{j=5/2,\, j_z=+5/2}$ state and the unoccupied states.
At $U=5~\mathrm{eV}$, the \ce{Ce}-$4f$ electrons are localized and no longer contribute directly to the Fermi surface.

\begin{figure}[t]
  \centering
  \includegraphics[width=0.9\columnwidth]{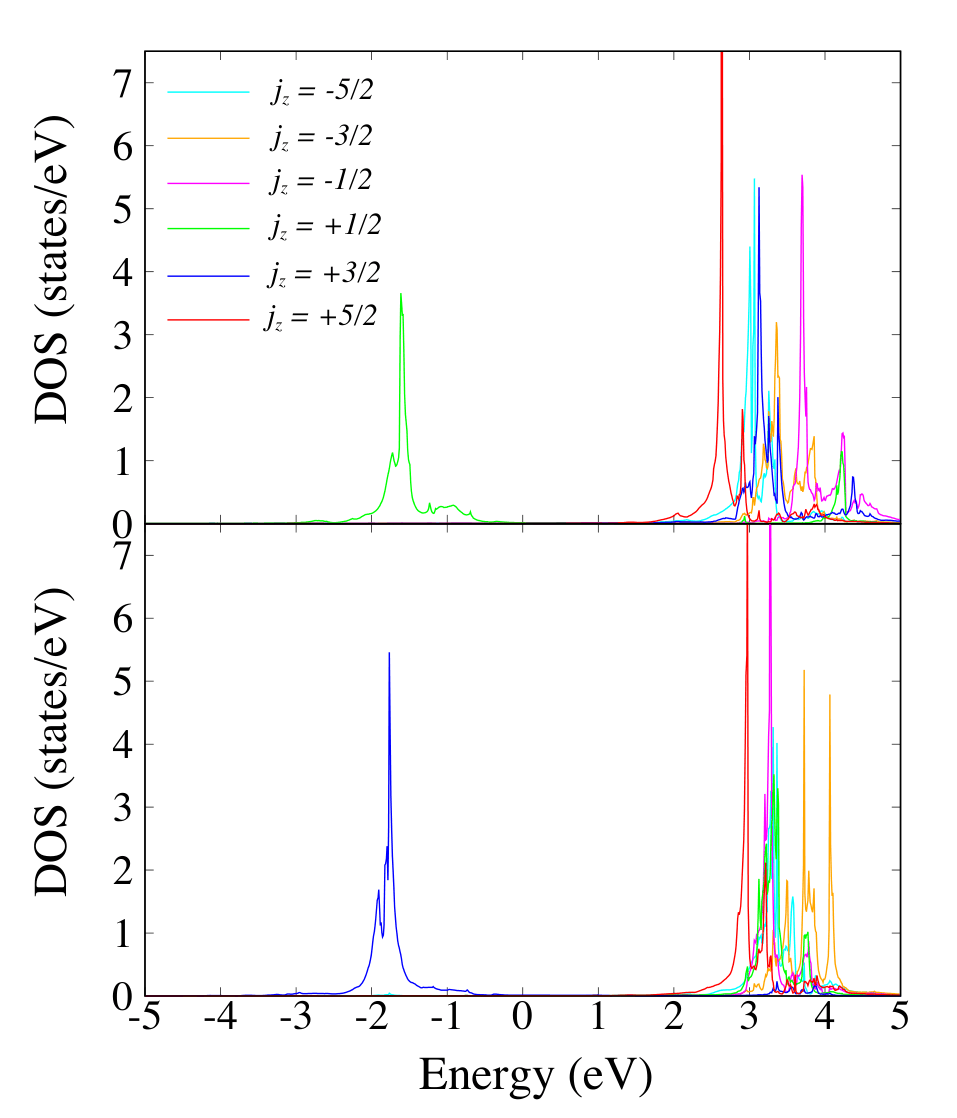}
  \caption{$j_z$-resolved DOS for each initially occupied state, $\ket{j=5/2,j_z=+1/2}$, $\ket{j=5/2,j_z=+3/2}$ and $\ket{j=5/2,j_z=+5/2}$, at $U=5~\mathrm{eV}$.
  }
  \label{fig:fig2}
\end{figure}

We can also obtain self-consistent solutions starting from an initial occupation corresponding to the $\ket{j=5/2,j_z=+1/2}$ and $\ket{j=5/2,j_z=+3/2}$ states.
These three representative occupied $4f$ states result from the combined effects of crystal-field splitting, spin-orbit coupling, and the breaking of time-reversal symmetry. 
Figure~\ref{fig:fig2} shows the $j_z$-resolved DOS at $U=5~\mathrm{eV}$ for $\ket{j=5/2,j_z=+1/2}$ and $\ket{j=5/2,j_z=+3/2}$ initial occupations, respectively.
In each case, the self-consistent solution retains the initially occupied $f$ state.
Together with the occupied $\ket{j=5/2,j_z=+5/2}$ case shown in Fig.~\ref{fig:fig1}, these results demonstrate that the three representative occupied $4f$ states can be obtained as self-consistent solutions.

\begin{figure}[h]
  \centering
  \includegraphics[width=1.0\columnwidth]{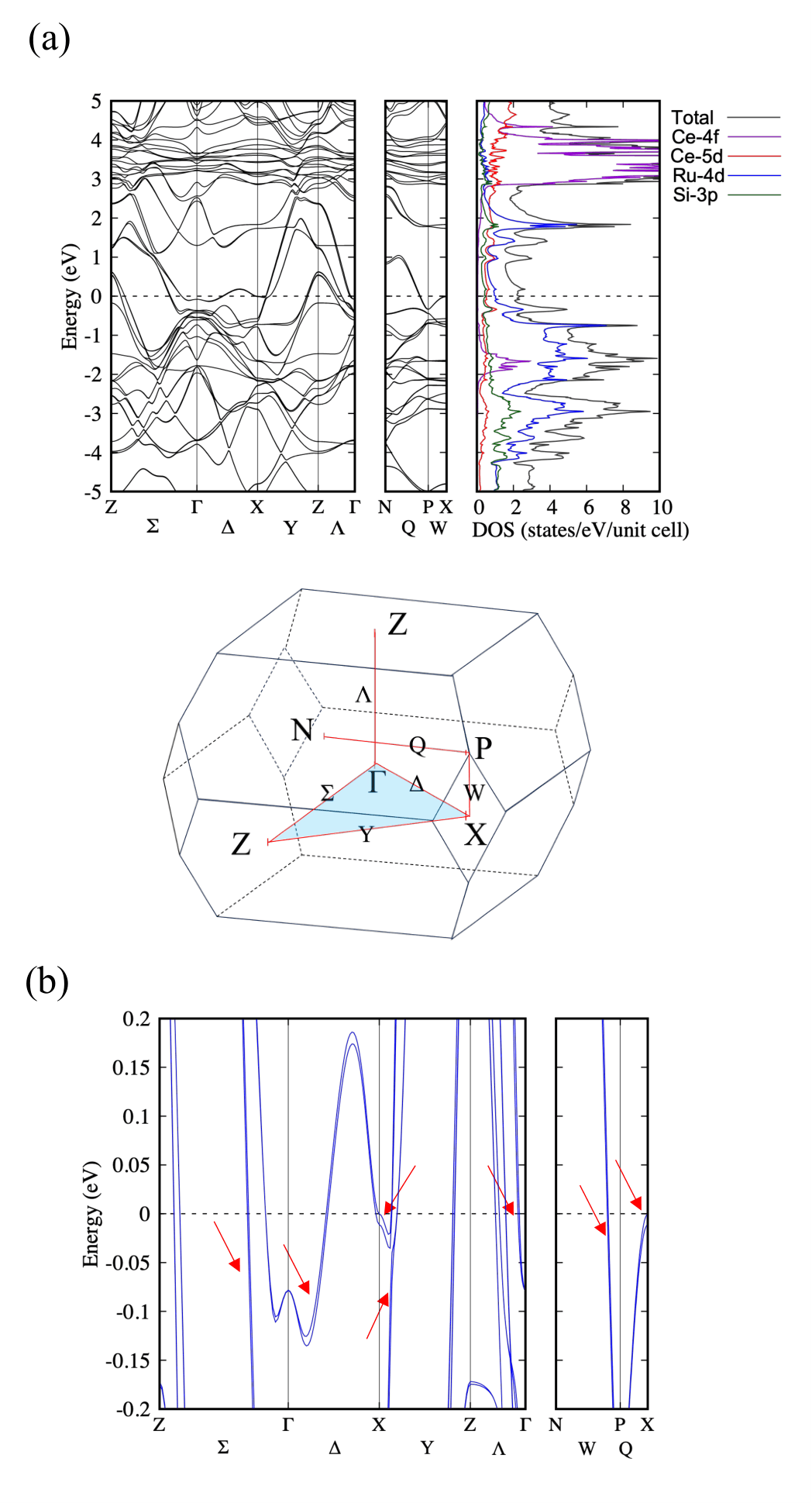}
  \caption{
  (a) Top: band structure of ferromagnetic \ce{CeRu2Si2} at $U=5~\mathrm{eV}$ (left) and atomic-orbital-resolved DOS (right).
      Bottom: first Brillouin zone of \ce{CeRu2Si2}.
  (b) Enlarged view of the band structure in top panel near the Fermi level. The red arrows highlight small exchange splitting of the conduction bands associated with the localized \ce{Ce}-$4f$ magnetism.}
  \label{fig:fig3}
\end{figure}

Although the \ce{Ce}-$4f$-states no longer contribute directly to the Fermi surface for large $U$, we find that their magnetic moments can still affect the electronic states near the Fermi level through exchange coupling with the conduction electrons.
To see this, we show the band structure and the atomic-orbital-resolved DOS of \ce{CeRu2Si2} with the occupied  $\ket{j=5/2,\, j_z=+5/2}$ state at $U=5~\mathrm{eV}$, in the upper panel of Fig.~\ref{fig:fig3}(a) with the Brillouin zone schematically shown in the lower panel.
The Fermi surface is formed mainly by bands derived from the \ce{Ce}-$5d$, \ce{Ru}-$4d$ and \ce{Si}-$3p$ orbitals.
The bands near the Fermi level are slightly split, as indicated by the red arrows in Fig.~\ref{fig:fig3}(b).
This suggests that the localized \ce{Ce}-$4f$ magnetism induces a small exchange splitting in the conduction bands.

\subsection{Intrinsic anomalous Hall effect}
\begin{figure}[t]
  \centering
  \includegraphics[width=\columnwidth]{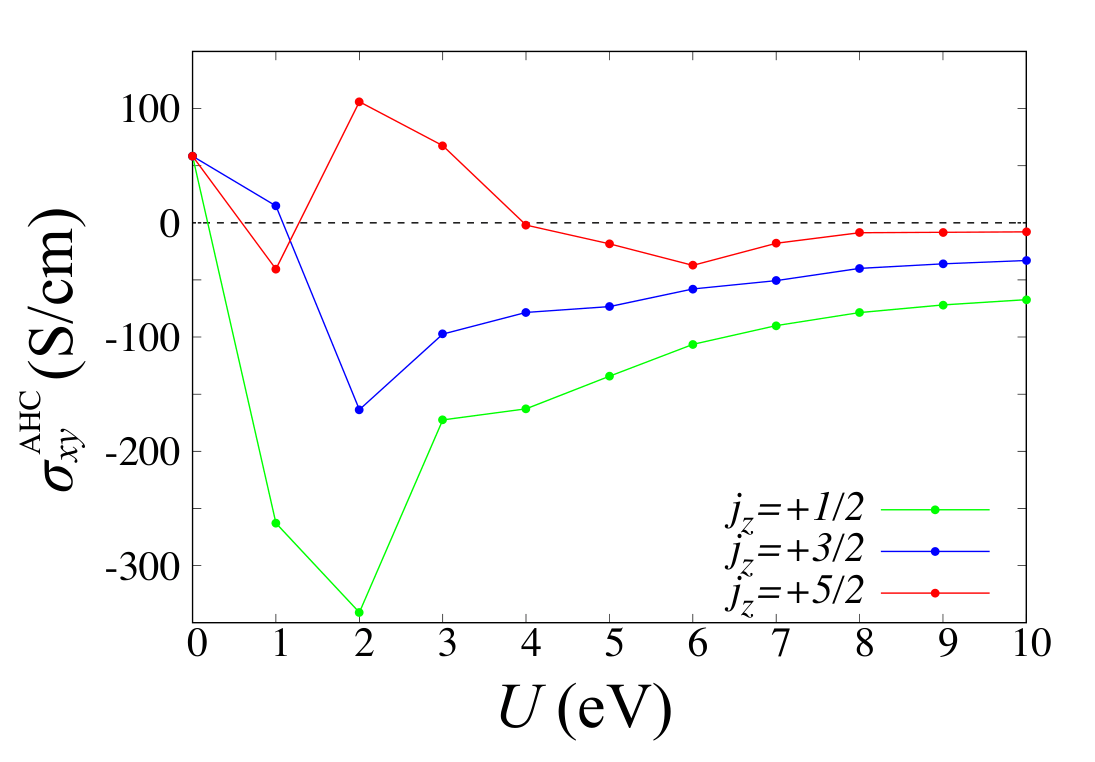}
  \caption{$U$-dependence of the AHC for different $\ket{5/2,j_z}$ occupied states.}
  \label{fig:fig4}
\end{figure}

We compare the AHC among the three occupied $\ket{j,j_z}=\ket{5/2,j_z}$ states with $j_z=+1/2$, $+3/2$, and $+5/2$.
Figure~\ref{fig:fig4} shows the $U$ dependence of the AHC for each occupied states. 
The AHC varies strongly in the small $U$ regime, while it approaches a finite value in the large-$U$ regime.
This behavior of the AHC is associated with the evolution of the \ce{Ce}-$4f$ electrons from itinerant to localized states with increasing $U$.
In the small-$U$ regime, which we refer to as the itinerant regime, the \ce{Ce}-$4f$ states remain near the Fermi level and directly contribute to the electronic states near the Fermi level. Consequently, changes in $U$ substantially modify the band structure near the Fermi level, leading to strong variations in the AHC.
Interestingly, the AHC remains finite for $U\geq 5~\mathrm{eV}$, where the $f$ state responsible for magnetism makes no direct contribution to the Fermi surface.
At $U=10~\mathrm{eV}$, the AHC reaches $-67~\mathrm{S/cm}$ for $j_z=+1/2$, $-33~\mathrm{S/cm}$ for $j_z=+3/2$, and $-8~\mathrm{S/cm}$ for $j_z=+5/2$, respectively.

\begin{figure}[t]
  \centering
  \includegraphics[width=\columnwidth]{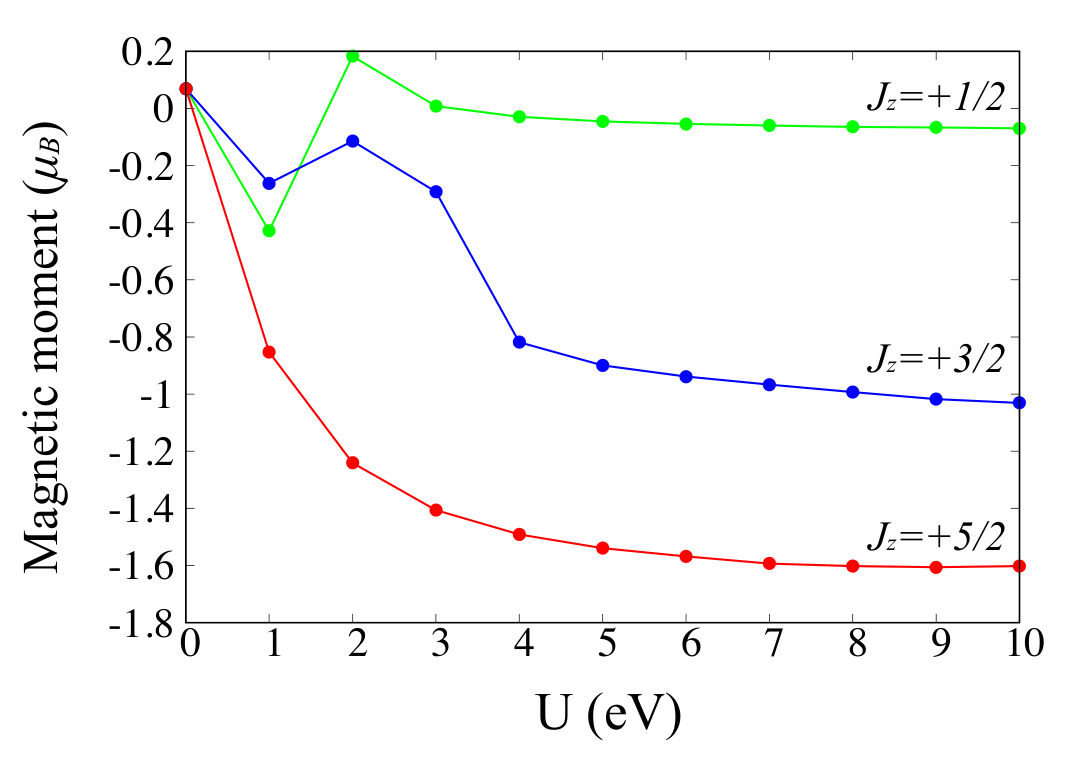}
  \caption{$U$-dependence of the magnetic moment of \ce{Ce}-$4f$ electron for different occupied $\ket{5/2,j_z}$ states.}
  \label{fig:fig5}
\end{figure}

The persistence of a finite AHC suggests that localized \ce{Ce}-$4f$ magnetism continues to affect the conduction electrons even after the \ce{Ce}-$4f$ states no longer contribute directly to the Fermi surface.
The dependence of the AHC on the occupied $\ket{j=5/2,j_z}$ state further indicates that the localized \ce{Ce}-$4f$ state plays an important role in determining the AHC.
Figure~\ref{fig:fig5} shows the $U$-dependence of magnetic moment of \ce{Ce}-$4f$ electron for different occupied $\ket{5/2,j_z}$ states. 
The magnetic moment is calculated by $\mu_z=-\mu_B(\expval{L_z}+2\expval{S_z})$ using the converged density matrix $n^{il}_{mm'ss'}$, where $L_z$ and $S_z$ denote the $z$ components of the orbital and spin angular momentum operators, respectively.
Here, $\mu_B$ denotes the Bohr magneton.
For large $U$, the magnetic moment of \ce{Ce}-$4f$-electron asymptotically approaches a constant value for each occupied $4f$-state.
As shown in Figs.~\ref{fig:fig4} and \ref{fig:fig5}, both the AHC and the \ce{Ce}-$4f$ magnetic moment approach state-dependent constant values in the large-$U$ regime.
This behavior suggests that the localized \ce{Ce}-$4f$ magnetic state plays an important role in determining the intrinsic AHC, even when the \ce{Ce}-$4f$ states do not contribute directly to the Fermi surface.

\begin{figure*}[h]
  \centering
  \includegraphics[width=15cm]{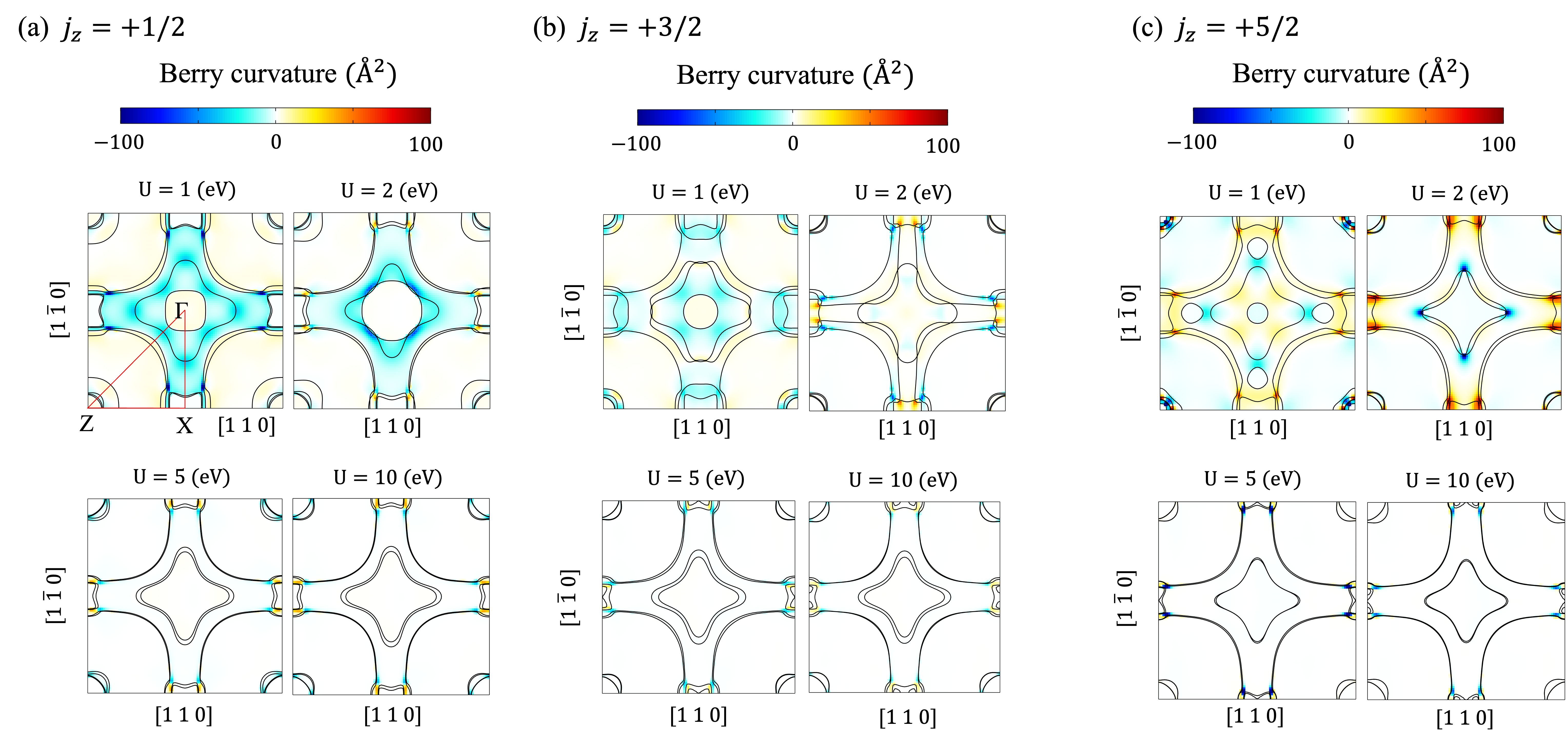}
  \caption{Berry curvature summed over the occupied bands on the $k_z=0$ plane for different occupied $4f$ states and values of $U$. Black curves indicate the Fermi surface. 
  (a) $\ket{j=+5/2,j_z=+1/2}$. (b) $\ket{j=+5/2,j_z=3/2}$. (c) $\ket{j=+5/2,j_z=+5/2}$.}
  \label{fig:fig6}
\end{figure*}

Finally, we analyze the Berry curvature summed over the occupied states, $\sum_{b}\Omega^z_{\bm kb}f(\varepsilon_{\bm kb}-\mu)$.
Figure~\ref{fig:fig6} shows the summed Berry curvature in the $k_z=0$ plane for the occupied $\ket{j=5/2,j_z=+1/2}$, $\ket{j=5/2,j_z=+3/2}$, and $\ket{j=+5/2,j_z=+5/2}$ states.
The region surrounded by the red-outlined triangle in Fig.~\ref{fig:fig6} (a) corresponds to the light-blue-filled triangle shown in the lower panel of Fig.\ref{fig:fig3}(a).
The black line represents the Fermi surface.
In the small-$U$ regime ($U=1,2~\mathrm{eV}$), the summed Berry curvature is broadly distributed over the Brillouin zone for all three $4f$ occupation states. 
Changes in $U$ within this regime significantly modify both the Fermi surface and the the Berry curvature distribution, resulting in strong variations in the AHC.
In the large-$U$ regime ($U=5$ and $10~\mathrm{eV}$), the Berry curvature becomes concentrated in limited regions of the Brillouin zone, while the Fermi surface remains nearly unchanged.
This is because the \ce{Ce}-$4f$ states are already located sufficiently far from the Fermi level.
Although the \ce{Ce}-$4f$ states no longer contribute directly to the Fermi surface in the large-$U$ regime, Fig. \ref{fig:fig3} shows that the \ce{Ce}-$4f$ magnetism induces a small exchange splitting of the conduction bands near the Fermi level. 
The resulting Berry-curvature distribution depends on which \ce{Ce}-$4f$ state is occupied, leading to quantitatively different AHCs.
These results indicate that the localized $\ce{Ce}$\MT{-}$4f$ magnetic state plays an crucial role in determining the magnitude of the intrinsic AHC.

\section{Conclusion}
In this study, we investigated the role of localized $f$ electrons in the intrinsic AHE by analyzing the AHC of ferromagnetic \ce{CeRu2Si2} for different occupied \ce{Ce}-$4f$ states.
We calculated the electronic structure using first-principles calculations within the LDA+$U$ approximation and evaluated the intrinsic AHC using the Wannier interpolation method.
We obtained the three self-consistent solutions with different localized $4f$ states by performing calculations starting from initial occupations corresponding to the $\ket{j=5/2,j_z=+1/2}$, $\ket{j=5/2,j_z=+3/2}$, and $\ket{j=5/2,j_z=+5/2}$ states.
We found that the AHC remains finite in the large-$U$ regime, where the \ce{Ce}-$4f$ states are localized and do not contribute directly to the Fermi surface, and that its magnitude depends on which \ce{Ce}-$4f$ state is occupied.
Our analysis shows that localized \ce{Ce}-$4f$ magnetism induces a small exchange splitting in the conduction bands near the Fermi level, producing Berry-curvature distributions that depend on the occupied \ce{Ce}-$4f$ state and hence different AHCs.
These results show that conduction bands derived from nonmagnetic orbitals can contribute to the intrinsic AHE through exchange splitting induced by localized $f$-electron magnetism.
This suggests that a large intrinsic AHE may be realized in f-electron materials when exchange splitting induced by localized 4f magnetism generates substantial Berry curvature near the Fermi level in conduction bands derived from nonmagnetic orbitals.
Recently, the microscopic origin of the intrinsic AHE in a localized-$f$ electron system has been investigated using dynamical mean field theory, highlighting the importance of spin-dependent hopping arising from nonlocal hybridization between conduction- and $4f$ electrons~\cite{kirikoshi2026nonlocalkondoexchangedrivenintrinsicanomalous}. 
Our first-principles results further show that the effect of localized $4f$ magnetism on the conduction bands depends strongly on the character of the occupied $4f$ state, producing distinct exchange splittings and Berry-curvature distributions near the Fermi level and, consequently, different intrinsic AHCs. These results establish the character of the occupied localized $4f$ state as an important factor controlling the Berry-curvature distribution and intrinsic AHC.

\begin{acknowledgments}
We acknowledge the use of the supercomputing system,
MASAMUNE-IMR, at CCMS, IMR, Tohoku University in
Japan. We are also grateful to K. Tanno for the technical supports.
This work was supported by JSPS KAKENHI Grants No. JP23K25827 (MTS), JP24K00581 (MTS), JP24K00588 (MTS), JP25K00947 (MTS), JP25K21684 (MTS), 26K17065 (IT).
\end{acknowledgments}

\newpage
\bibliography{samp_capitalization_protected.bib}

\end{document}